\documentclass[twocolumn,prl,aps,superscriptaddress,showpacs,amsmath,amssymb,floatfix]{revtex4-1}
\usepackage{graphicx}
\usepackage{amssymb}
\usepackage{amsmath}
\usepackage{epsfig}
\usepackage{color}
\usepackage{tabu}
\usepackage{mathtools}
\usepackage[colorlinks,linkcolor=blue,anchorcolor=blue,citecolor=blue,urlcolor=blue]{hyperref}
\usepackage{physics}
\usepackage{float}
\usepackage{diagbox}

\begin{document}

\title{Nutation dynamics and multifrequency resonance in a many-body seesaw} 
\author{Hong-Ze Xu}
\affiliation{CAS Key Laboratory of Quantum Information, University of Science and Technology of China, Hefei, 230026, China}
\author{Shun-Yao Zhang}
\affiliation{CAS Key Laboratory of Quantum Information, University of Science and Technology of China, Hefei, 230026, China}
\author{Yu-Kai Lu}
\thanks{Present address: Department of Electrical Engineering, Princeton University, Princeton, New Jersey 08544, USA}
\affiliation{CAS Key Laboratory of Quantum Information, University of Science and Technology of China, Hefei, 230026, China}
\author{Guang-Can Guo}
\affiliation{CAS Key Laboratory of Quantum Information, University of Science and Technology of China, Hefei, 230026, China}
\affiliation{Synergetic Innovation Center of Quantum Information and Quantum Physics, University of Science and Technology of China, Hefei, Anhui 230026, China}
\affiliation{CAS Center For Excellence in Quantum Information and Quantum Physics, University of Science and Technology of China, Hefei, Anhui 230026, China}
\author{Ming Gong}
\email{gongm@ustc.edu.cn}
\affiliation{CAS Key Laboratory of Quantum Information, University of Science and Technology of China, Hefei, 230026, China}
\affiliation{Synergetic Innovation Center of Quantum Information and Quantum Physics, University of Science and Technology of China, Hefei, Anhui 230026, China}
\affiliation{CAS Center For Excellence in Quantum Information and Quantum Physics, University of Science and Technology of China, Hefei, Anhui 230026, China}
\date{\today }

\begin{abstract}
The multifrequency resonance has been widely explored in the context of single-particle models, of which the modulating Rabi model has been the most widely investigated. It has been found that with diagonal periodic modulation, steady dynamics can be realized in some well-defined discrete frequencies. These frequencies are independent of off-diagonal couplings. In this work, we generalize this physics to the many-body seesaw realized using the tilted Bose–Hubbard model. We find that the wave function will recover to its initial condition when the modulation frequency is commensurate with the initial energy level spacing between
the ground and the first excited levels. The period is determined by the driving frequency and commensurate ratio. In this case, the wave function will be almost exclusively restricted to the lowest two instantaneous energy levels. By projecting the wave function to these two relevant states, the dynamics is exactly the same as that for the spin precession dynamics and nutation dynamics around an oscillating axis. We map out the corresponding phase diagram, and show that, in the low-frequency regime, the state is thermalized, and in the strong modulation limit,
the dynamics is determined by the effective Floquet Hamiltonian. The measurement of these dynamics from the mean position and mean momentum in phase space are also discussed. Our results provide new insights into multifrequency resonance in the many-body system.
\end{abstract}
\maketitle

The multifrequency resonance has been widely explored in some of the single-particle models \cite{haroche1970modified, 
PhysRevA.80.050302,PhysRevB.81.035205}, in which the two-level Rabi model subjects to diagonal modulation was most widely 
investigated \cite{JOUR2016,PhysRevA.96.033802,PhysRevLett.87.246601,Longhi_2006}. The simplest model for this mechanism
can be written as \cite{JOUR2016,PhysRevA.75.063414,Oliver1653}
\begin{equation}
H = -\frac{\Delta}{2} \sigma_x - \frac{\varepsilon(t)}{2} \sigma_z, 
\end{equation}
where $\Delta$ is the coupling strength between the ground state and excited state, $\varepsilon(t)$ is the time-dependent 
bias, and $\sigma_x$ and $\sigma_z$ are Pauli matrices. We can choose 
\begin{equation}
	\varepsilon(t) = \epsilon_0 + A \cos(\omega t),
\end{equation}
where $A$ is the driving amplitude, $\varepsilon_0$ is the static detuning and $\omega$ is the modulation frequency. 
The Schr\"{o}dinger equation for $H$ can be written as $i \partial_t |\psi\rangle = H |\psi\rangle$. We can choose a 
rotating operator as
\begin{equation}
	U(t) = \exp[{i \over 2} (\varepsilon_0 t + {A \over \omega} \sin(\omega t)) \sigma_z],
\end{equation}
and write the wave function as $|\psi\rangle = U(t) |\psi_\text{r}\rangle$, where $|\psi_\text{r}\rangle$ is the wave 
function in the rotating frame. We find that $ i\partial_t |\psi_\text{r}\rangle = H^\text{r} |\psi_\text{r}\rangle$, 
with 
\begin{equation}
\begin{aligned}
 H^\text{r} = U^\dagger(t) H U(t) - i U^\dagger(t) \frac{d U(t)}{dt} 
	=-\frac{\Delta}{2}
   \left(\begin{array}{cc} 
	0 & Q \\
	Q^* & 0
\end{array}\right),
\end{aligned}
\end{equation}
where $Q = \sum_{n=-\infty}^{\infty} J_{n}\left(\frac{A}{\omega}\right) e^{-i\left(n \omega+\varepsilon_{0}\right) t}$, with 
$J_{n}(x)$ being the Bessel function of the first kind. Steady dynamics can be realized when $n \omega+\varepsilon_{0} 
= 0$ for some integer $n$, which yields multifrequency resonance. This resonance is independent of the off-diagonal
coupling $\Delta$. Since the above Rabi model can be readily realized using some of the quantum simulators, this physics has been
implemented in superconducting qubits \cite{PhysRevLett.96.187002,PhysRevLett.101.017003,PhysRevB.81.024520,PhysRevB.83.180507}, ultracold atoms \cite{PhysRevA.95.013827}, and NV color centers \cite{PhysRevA.82.033839,Jiang267,Neumann1326,PhysRevLett.111.067601}. An 
intensive review of this physics in the Rabi model can be found in Ref. \cite{Xie_2017}. This method has also been applied 
in medicine for magnetic resonance imaging \cite{Eles2010Two, Victor2020multiphoton}. Moreover, the similar physics can also 
be found in the other models without these energy levels, such as the nonlinear oscillator \cite{PhysRevLett.48.714,PhysRevA.34.726,PhysRevLett.83.3406,PhysRevB.71.140508}, nanomechanical resonators \cite{PhysRevLett.94.156403,PhysRevLett.99.040404}, and Josephson junctions \cite{PhysRevB.68.060504,PhysRevLett.93.207002}. However, the physics in the many-body system, which involves much more complicated energy levels, is rarely 
discussed.

\begin{figure}
\centering
\includegraphics[width=0.48\textwidth]{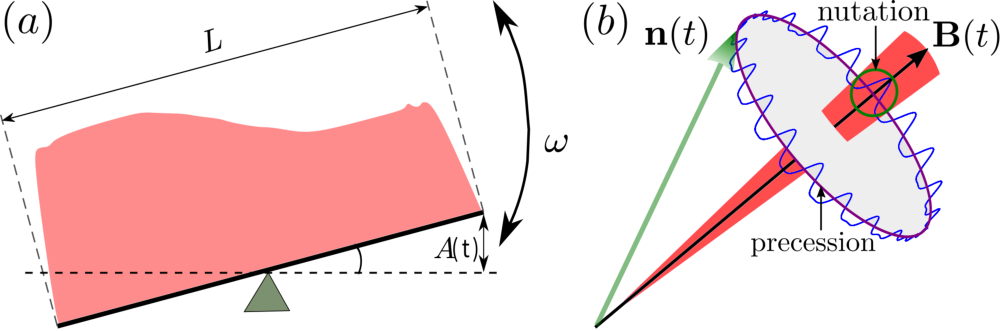}
\caption{(a) Quantum seesaw model constructed by a finite chain with length $L$ under hard-wall boundary condition. The modulation frequency 
$\omega$, inclination $\alpha$ are controllable in experiments. The change of potential at the two ends is $A(t) = \alpha \sin(\omega t)\frac{L}{2}$. (b) Nutation and precession dynamics for a spin about an modulating magnetic field. The corresponding spin vector is defined as ${\bf n}(t) = \langle \psi(t)|\boldsymbol{\sigma}|\psi(0) \rangle$, where $\boldsymbol{\sigma}$ are Pauli matrices. Here the smaller, faster, periodic oscillation is called the nutation dynamics, while the slower one is called the precession dynamics. }
	\label{fig-fig1}
\end{figure}
 
In this work, we generalize this concept to the realm of many-body physics based on a modulating quantum seesaw realized by the
Bose-Hubbard model, in which some new dynamics beyond the above single-particle picture can be realized. We find that: (I) 
When the modulation frequency is commensurate with the level spacing between the ground state and the first excited state of the
Hamiltonian at time $t = 0$, periodic recovery of the wave function can be realized, with a period determined by the tunable 
commensurate ratio $\beta$. (II) By projecting the wave function to the two lowest states, the many-body dynamics is reduced to 
a spin model 
precession about an oscillating magnetic field, which yields a new mechanism to the realization of the multifrequency resonance 
via nutation dynamics \cite{driben2016precession, ciornei2011magnetization}. This 
picture even has a single-particle and classical analogous. 
(III) This phase can be realized only when the commensurate ratio is large enough. In the low modulation limit beyond 
the adiabatic limit, the many-body state will quickly approaches the thermalized state. In the high-frequency limit, this 
nutation dynamics is suppressed, and the dynamics is determined by the Floquet Hamiltonian. We have also discussed the 
experimental detection of these phases and discuss their stability with non-integrability interactions. Our 
results are stimulating for the exploring of the multifrequency resonance in the many-body system.

{\it Physical model and dynamics}. We consider the following driven Bose-Hubbard model in a finite chain
\begin{eqnarray}
	H && = -J\sum_{i=1}^{L-1} (b_i^\dagger b_{i+1} + \text{h.c.}) + \frac{U}{2}\sum_{i=1}^L n_i(n_i-1)  \nonumber \\
	  && + \alpha \sum_{i=1}^L \sin(\omega t) (i-{L+1 \over 2})  n_i,
	  \label{eq-H}
\end{eqnarray}
which is schematically shown in Fig. \ref{fig-fig1}. Here $b_i^\dagger$ $(b_i)$ are the creation (annihilation)
operators at the $i$-th lattice site, $n_i=b_i^\dagger b_i$ is the number operator, $J$ is the tunneling strength and 
$U > 0$ is the on-site many-body interaction. 
In following, we set $J = 1$ as the basic energy scale. In experiments, $J\sim 2\pi \times \mathcal{O}(0.1)$ kHz \cite{dai2017four,
trotzky2012probing,bukov2015prethermal, pigneur2018relaxation}. The last term represents the quantum seesaw, which can be 
realized by a tilted magnetic field with modulation frequency $\omega$ and inclination $\alpha$. This tilted potential has 
been realized in experiments \cite{simon2011quantum,preiss2015strongly,geiger2018observation,
kennedy2015observation}. The hard wall boundary condition has been realized with 
a box potential in several groups, with $L$ typically from 10 to 100 \cite{lopes2017quantum,gaunt2013bose,mukherjee2017homogeneous,eigen2016observation}. 
The data we will present are obtained by exact diagonalization (ED) and time-evolving block decimation (TEBD) methods \cite{vidal2003g,vidal2004efficient}. In the simulation, we choose
$\omega = \beta \Delta E_{12}$, where $\Delta E_{12} = E_2-E_1$ is the level spacing between the ground state and first excited state of $H(t=0)$ and $\beta$ is a rational number termed as the commensurate ratio.

\begin{figure}
\centering
\includegraphics[width=0.45\textwidth]{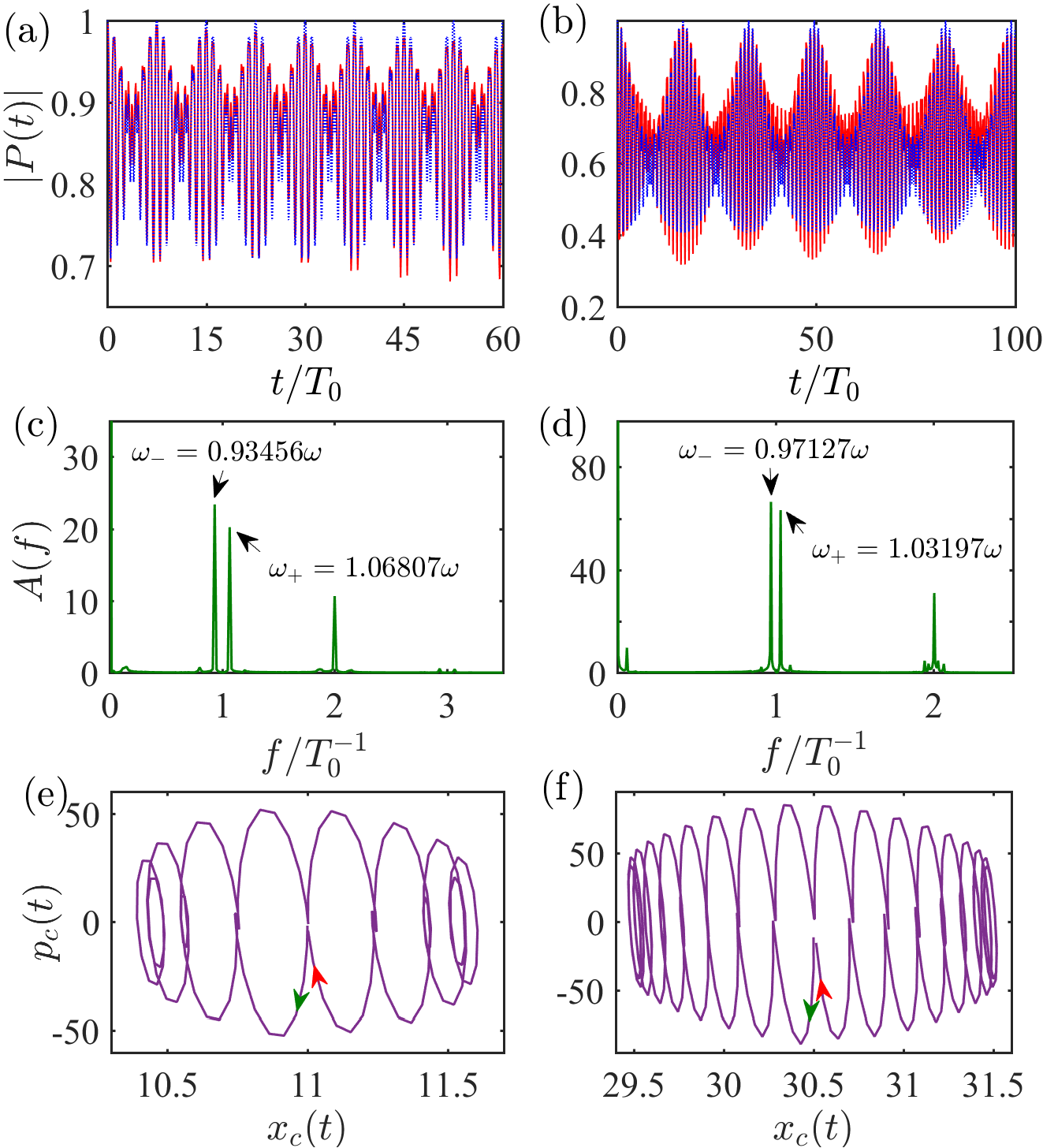}
\caption{(a) - (b) Periodically recovery of wave function. (a) Red line for $L=21$, $\beta=15$, $N=10$, $U=\sqrt{2}$, $\alpha=1/3$ and blue dotted line is given by Eq. \ref{eq-first-order-two-level}, for $g=1.27$, $A=\lambda=0$, $\epsilon=\Delta E_{12}/2$. (b) Red line for $L=60$, $\beta=33$, $N=35$, $U=0.8$, $\alpha=0.05$ and blue dotted line for $g=1.83$, $A=\lambda=0$, $\epsilon=\Delta E_{12}/2$.  (c)-(d) Fourier spectroscopy of $P(t)$. The arrows correspond to $\omega_{\pm} = \omega\pm \omega'$. 
(e)-(f) Trajectory in phase space by mean position and mean momentum. The arrows indicate the direction of the time evolution.}
\label{fig-fig2}
\end{figure}
 
\begin{figure}
\centering
\includegraphics[width=0.48\textwidth]{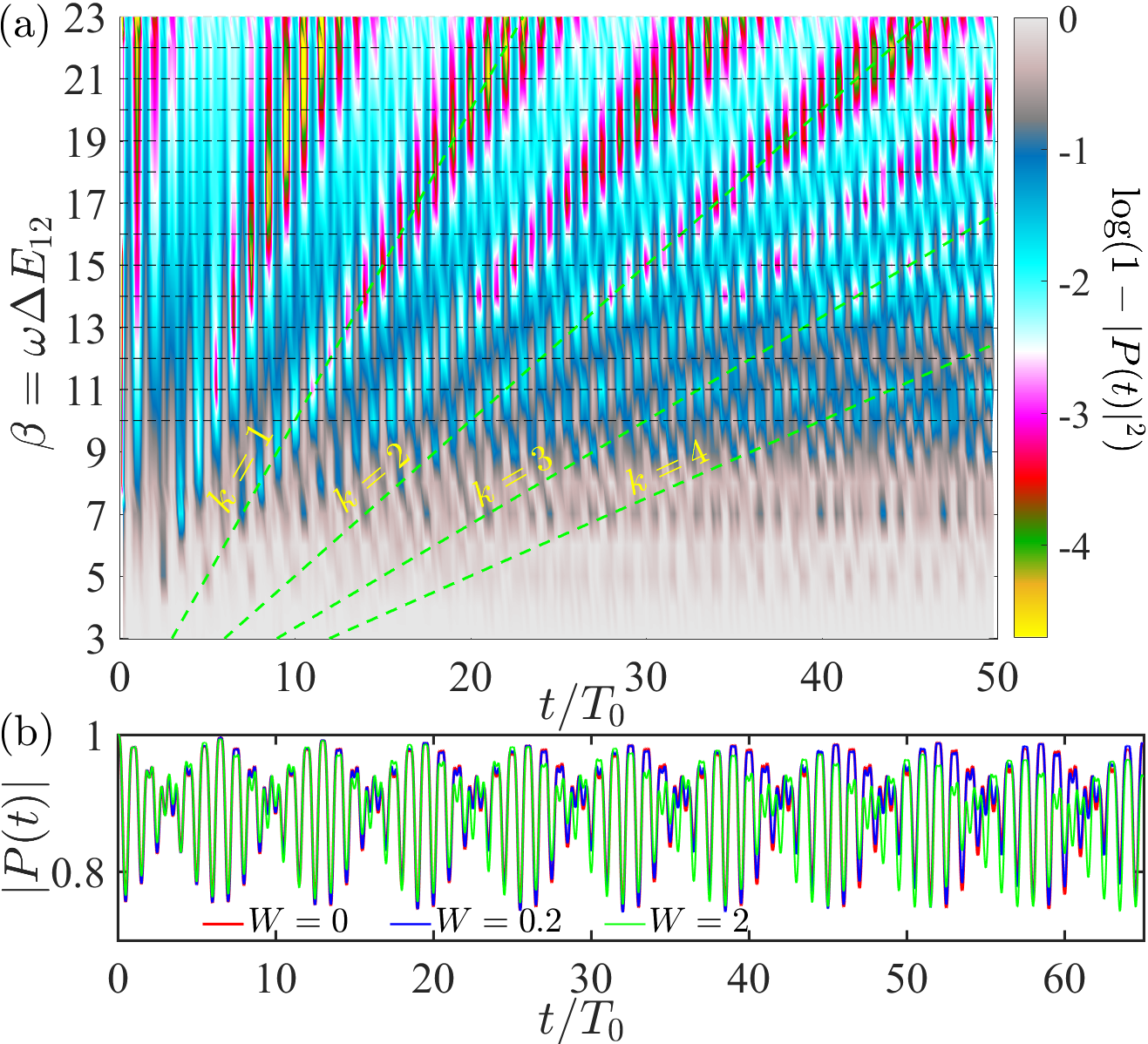}
	\caption{ (a). The probability $|P(t)|^2$ as a function of $\beta$ and $t/T_0$ with $L=21$, $N=10$, $U=\sqrt{2}$, $\alpha=1/(3\sqrt{5})$. The green dashed lines are given by $t/T_0=k\beta$, where $k=1$, 2 , 3, $\cdots$.  (b). The influence of disorder in the multifrequency resonance phase with $L=12$, $N=5$, $U=0.8$, $\alpha=0.3$. For different disorder strength, we fixed driving frequency $\omega=13\Delta E_{12}$, where the $\Delta E_{12}$ is given by $W=0$ and averaged by 10 realizations. }
\label{fig-fig3}
\end{figure} 

Let us first consider the dynamics in a short chain ($L = 21$, $\beta = 15$) and a long chain ($L = 60$, $\beta = 33$) in Fig. \ref{fig-fig2}. The same physics can be found by choosing some other parameters. To measure whether the wave function 
will recover to its initial state $|\psi(0)\rangle$ (the ground state of $H(t=0)$), we measure the overlap between them via $P(t) = \langle \psi(0)|\psi(t)\rangle$ \cite{heyl2013dynamical,jurcevic2017direct} (see Fig. \ref{fig-fig2}(a) - (b)), which 
is related to the Ramsey interferometry in experiments \cite{goold2011orthogonality,knap2012time,cetina2016ultrafast}. 
To determine the period, we have also calculated the Fourier transition of this recovery probability (see Fig. \ref{fig-fig2}(c) - (d)). In both cases, the wave function will 
almost recover to its initial state with $|P| > 93\%$. This period can be precisely determined by the Fourier spectra in the time domain with two
frequencies $\omega_{\pm} = \omega \pm \omega^{\prime}$, where $\omega$ is the driving frequency of the seesaw. We find that $T = 2\pi/\omega^{\prime} = \beta T_0$, 
with $T_0 = 2\pi/\omega$ being the period of the driving field. This dynamics can persist for an extraordinarily long time.
Here we only consider the case that $\beta$ to be integer values for simplicity, and a more general proof will be presented below, 
from which our conclusion is also true for rational numbers.

To detect this dynamics, we investigate the mean position $x_c$ and mean momentum $p_c$ in the phase space in Fig. \ref{fig-fig2}(e) - (f), by 
defining $x_c(t) = \sum_m \langle \psi(t)| b_m^\dagger b_m |\psi(t)\rangle m/N$, where $N$ is the total number of particles and $p_c(t)= \text{d}x_c(t)/\text{d}t$ (we have used $\text{d}t = 0.001$ in the calculation of $p_c$). In the
phase space, these two variables construct an almost closed trajectory after one period $T$. For the two sets of parameters in Fig. \ref{fig-fig2}, the real space displacement is roughly one or two lattice sites, and the change of mean momentum is slightly larger (in the standard unit of momentum) than this value. These two variables can be measured in both real and momentum spaces from the time-of-flight spectroscopy \cite{wang2012spin,hart2015observation,atala2013direct,anderlini2007controlled}.

In order to more clearly see the relationship between the driving frequency and the response period, we plot the recovery probability 
in Fig. \ref{fig-fig3}(a). When $\beta$ is small, we find a regime with complicated dynamics; however, with the increasing of $\beta$,
the periodic modulation of $|P(t)|^2$ can be observed as a function of evolution time. We find that when $t/T_0 = k\beta$ $(k=1,2,3,
\cdots)$, the probability $|P(t)|^2$ will approaches a peak with the largest recovery $|P|\rightarrow 1$. 
That means the response period $T = \beta T_0$. 

{\it The influence of disorder}. Next, we consider the influence of on-sites disorder for the multifrequency resonance phase. 
We introduce the disordered term $H_D = \sum_i^L h_i n_i$ in Eq. \ref{eq-H}, where $h_i\in[-W,W]$ and $W$ is the disorder 
strength. This is a typical term to break the integrability of the Bose-Hubbard model. 
We choose $L=12$, $N=5$, $U=0.8$, $\alpha=0.3$ with fixed frequency $\omega=13\Delta E_{12}$, where the 
$\Delta E_{12}$ is the gap of Eq. \ref{eq-H} at $t=0$. The results are shown in Fig. \ref{fig-fig3}(b) averaged by 10 realizations. 
In our model, the recovery period is strongly dependents on the energy gap of the system parameters at the initial moment. 
Therefore, for weak disorder strength, $W=0.2$,  we can see that the curve almost coincides with $W = 0$, because the gap is 
almost unchanged. However, for strong disorder strength, $W=2$, the gap will be changed, so, the curve is significantly different 
from $W=0$ at long time. However, the periodical recovery of the probability can still be seen clearly, indicating that the
dynamics is robust against disorder and the related non-integrability interactions.

\begin{figure}
\centering
\includegraphics[width=0.45\textwidth]{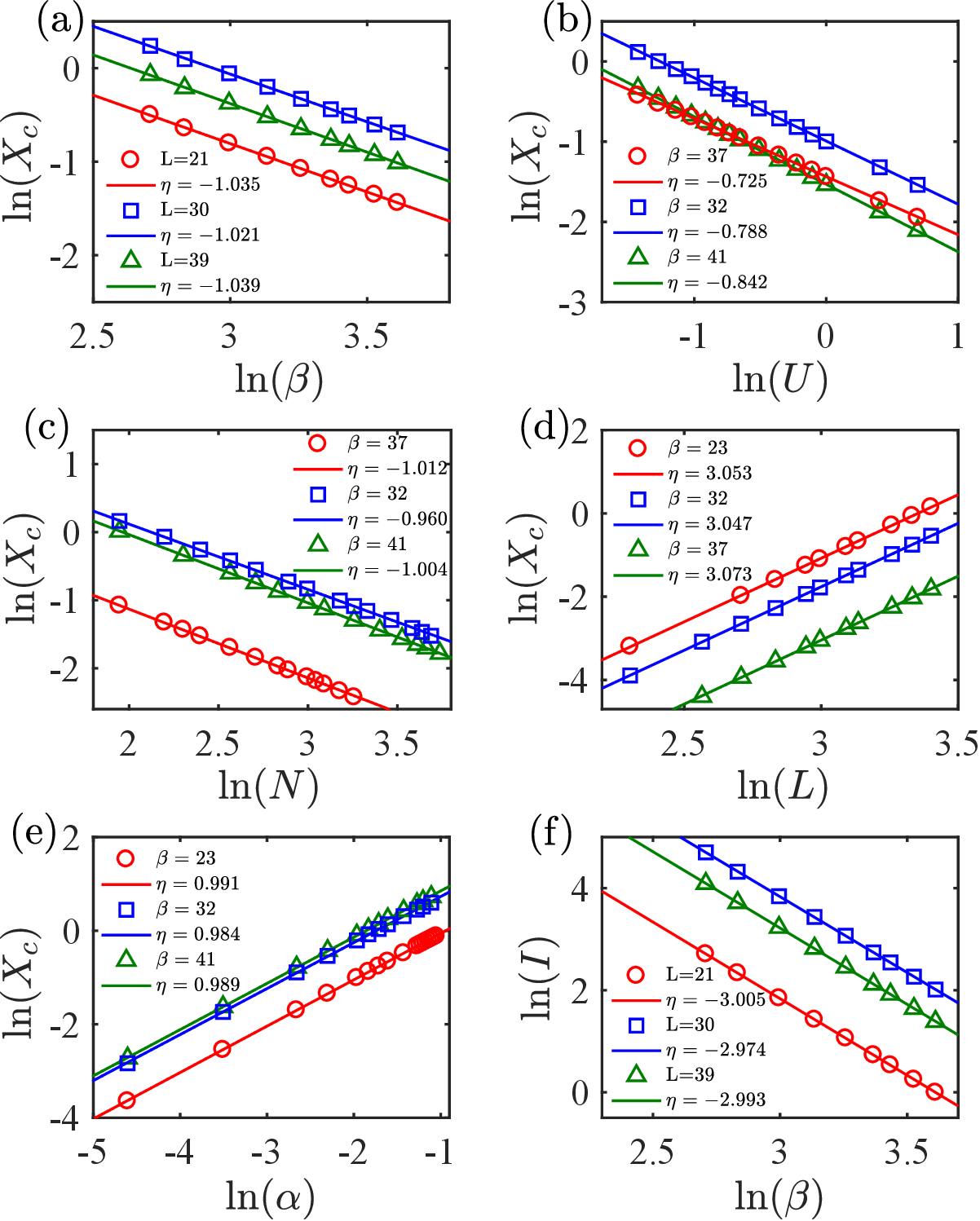}
\caption{(a) - (e) The maximal displacement $X_c$ (see Eq. \ref{eq-Xc}) as a function of $\beta$, $U$, $N$, $L$ and $\alpha$. (f) The area enclosed by the trajectory in phase space as a function of $\beta$. Parameters for open symbols \begin{scriptsize}\textcolor[rgb]{1,0,0}{$\bigcirc$}\end{scriptsize} are $L=21$, $N=10$, $U=\sqrt{2}$, $\alpha=1/(3\sqrt{5})$;  \textcolor[rgb]{0,0,1}{$\square$} are $L=30$, $N=15$, $U=0.8$, $\alpha=0.1$ and  \textcolor[rgb]{0,0.5,0}{$\bigtriangleup$} are $L=39$, $N=22$, $U=\sqrt{3}/2$, $\alpha=0.05$.}
\label{fig-fig4}
\end{figure}

{\it The influence of parameters on the measurement}. We next explore how the parameters influence the mean position displacement, by defining
\begin{equation}
	X_c = \frac{\text{max}(x_c) - \text{min}(x_c)}{2}  \sim s^{\eta},
	\label{eq-Xc}
\end{equation}
where $s$ may refer to $\beta$, $U$, $N$, $L$, $\alpha$ {\it etc.}.
The exponents for these five cases are $\eta \simeq -1.0, -0.8, -1.0$, $3.0$ and $1.0$, respectively; see Fig. \ref{fig-fig4}(a) - (e). This means, in experiments, the larger center of mass displacement can be found with relatively smaller modulation frequency, interaction strength and total number of particles, and relatively larger number of lattice site $L$ and inclination $\alpha$. We also measure the area of the trajectory enclosed by $p_c$ and $x_c$ using $I = \frac{1}{2\pi}\oint p_c \text{d} x_c \sim s^{\eta}$. This quality has a number of interesting features. In the adiabatic limit, it should be quantized (in the unit of Planck constant), which has played a fundamental role in history in the development of quantum theory and quantization condition in Sommerfeld theory. 
In Fig. \ref{fig-fig4}(f), we find that $I$ has 
the same scaling as Eq. \ref{eq-Xc} with exponent $\eta  \simeq -3.0$. This value is not quantized due to non-adiabatic evolution. 

\begin{figure}
\centering
\includegraphics[width=0.45\textwidth]{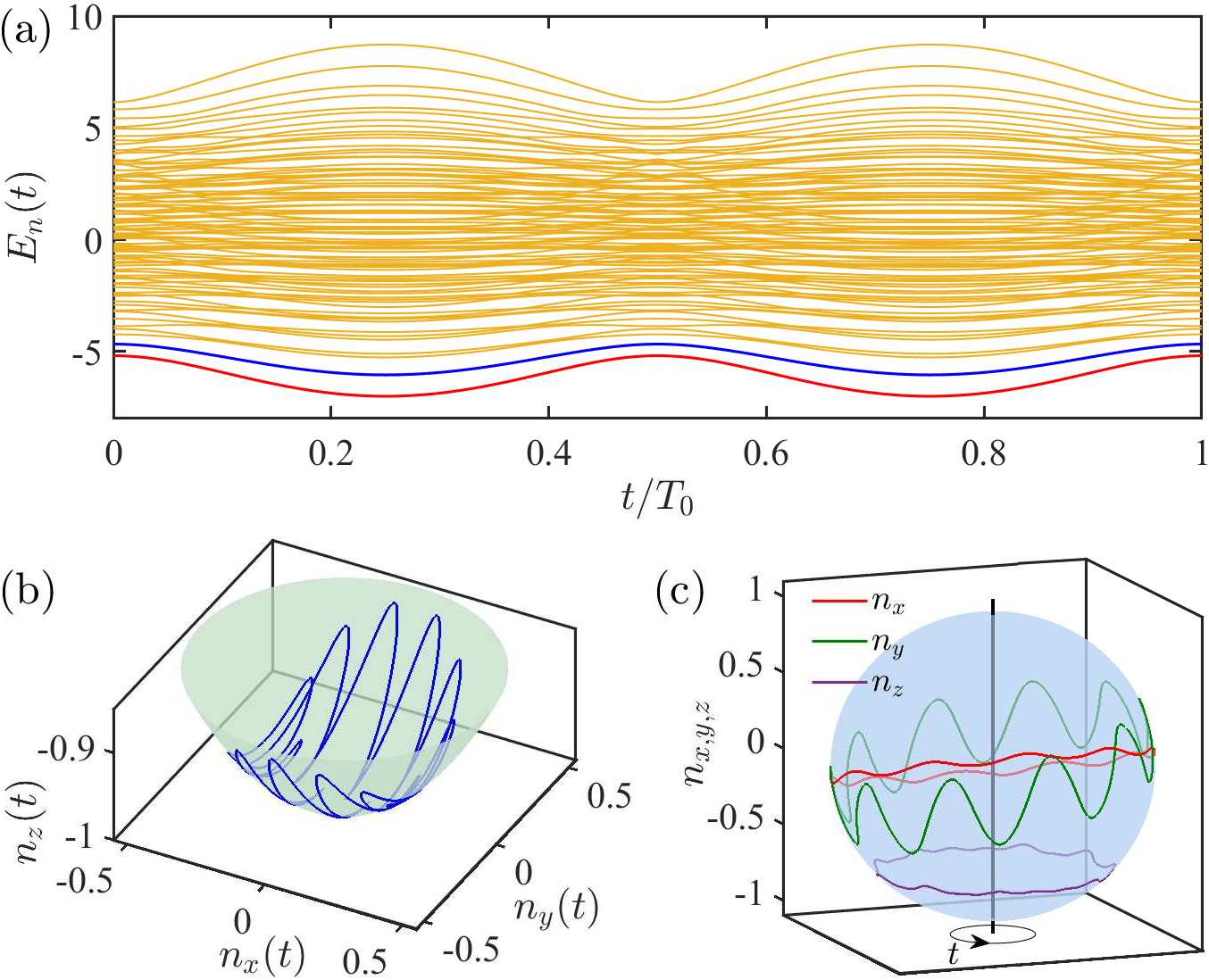}
	\caption{(a) Instantaneous eigenvalues for $L=7,N=3,U=0.8,\alpha=0.5,\omega=11\Delta E_{12}$ and 
	$\Delta E_{12} = 0.518804$. (b) The dynamics of the spin vector ${\bf n}$ in the Bloch sphere,
	with $A=\lambda=0,g=\sqrt{2},\epsilon=0.53,\omega=10\Delta E_{12}$. (c) The corresponding precession and nutation dynamics in the three directions, with the arrow marks the evolution in one full period.}
\label{fig-fig5}
\end{figure}
{\it Mapping to a spin vector about an oscillating magnetic field}. This model provides a new mechanism for the realization of multifrequency resonance with a controllable commensurate period, which will be termed as multifrequency resonance phase. To pin down the underlying mechanism for this dynamics, we 
now project the wave function $\psi(t)$ from the time-dependent Schr\"odinger equation to the instantaneous eigenstates $\phi_n$, where $H(t)\phi_n(t) = 
\epsilon_n(t) \phi_n(t)$ with $\epsilon_n$ to be arranged in increasing order. The idea is quite similar 
to the derivation of geometry phase in topological physics \cite{berry1985classical, berry2009transitionless, del2013shortcuts, aharonov1984topological}, but now we need to consider a few low-lying eigenstates as
\begin{equation}
	\psi(t) = c_1 \phi_1(t) + c_2 \phi_2(t) + \cdots.
	\label{eq-proj}
\end{equation}
We will focus on the regime when $\omega$ is smaller than the whole bandwidth (see the eigenvalues of $H(t)$ in Fig. \ref{fig-fig5}(a), with bandwidth $W_b = \text{max}(\epsilon_n) - \text{min}(\epsilon_n) = 11.33844$). We find that $|c_1|^2 + |c_2|^2$ is always larger than 0.9, indicating that almost all the wave functions, during the long-time evolution, is almost restricted to the lowest two instantaneous eigenstates. In this way, we can keep only these two major terms. Then $\psi(t)$ can be the solution of the  Schr\"odinger equation corresponding to the following equivalent Hamiltonian,
\begin{equation}
	\mathcal{H}_\text{eq}(t) =  \varepsilon (t)  + b^i(t) \sigma_i =  \varepsilon(t) + {\bf B}(t)\cdot \boldsymbol{\sigma},
	\label{eq-two-level}
\end{equation}
where the Pauli matrices act on the subspace constructed by the lowest two states in Eq. \ref{eq-proj} and ${\bf B} = (b^x, b^y, b^z)$. 
In this manner, we map the many-body model to the spin dynamics about an oscillating 
magnetic field ${\bf B}(t)$, in which the spin exhibits both (slower) precession and (faster) nutation dynamics (see Fig. \ref{fig-fig1}(b)). The nutation dynamics comes from the periodic modulation of the rotating axis. We also describe this dynamics on the Bloch sphere, showing in Fig. \ref{fig-fig5}(b) and the corresponding precession and nutation dynamics in the three directions are shown in Fig. \ref{fig-fig5}(c). This kind of dynamics is analogous to the dynamics in astrophysics \cite{herring1991forced,mathews2002modeling}, in which 
these two dynamics are prevailing. 

For the general case, we let ${\bf B}(t) = (y(t)+\lambda, 0, f(t)+\epsilon)$. Let us assume the initial state to be 
$ |\psi(0)\rangle =|\downarrow\rangle $ and to the leading order via perturbation theory we find
\begin{equation}
\left\{
\begin{aligned}
	& |\psi(t)\rangle =\frac{1}{\sqrt{|c_1(t)|^2 + |c_2(t)|^2}} (c_1(t)|\downarrow\rangle + c_2(t)|\uparrow\rangle),\\ 
	& c_1(t)=1-i \int_0^t f(t) dt, \quad c_2(t)= -i \int_0^t e^{i \Delta E_{12} t} y(t) dt,
	\label{eq-first-order-two-level}
\end{aligned}
  \right.
\end{equation}
where $\Delta E_{12}=2\sqrt{ \epsilon^2+ \lambda^2}$. If $f(t)$ and $y(t)$ are smooth functions with the same modulating frequency 
$\omega$, that means $f(t)$ and $y(t)$ have the same period $T_0 = 2\pi / \omega$. On the other hand, $e^{i\Delta E_{12}t}$ has the period $T^\prime = 2\pi / \Delta E_{12} $. So when we let $\omega = \frac{p}{q} \Delta E_{12} = \beta \Delta E_{12}$, where $p$ 
and $q$ are relatively prime numbers with $p > q$, we can get $T^\prime /T_0 = p/q$. That is to say the period of $e^{i \Delta E_{12} t} y(t)$ is $T = pT_0 = qT^\prime$ and $P(t)=\langle \psi(0)| \psi(t)\rangle $ also has the period $T$. However, the case with 
$\beta \in \mathbb{Z}$ has the simplest recovery curve. 

In our model, we find that the dynamics can be reproduced by letting $y(t) = g\sin \omega t$ and $f(t) = A\sin \omega t$. One can easily obtain $c_1(nT) = 1$ and $c_2(nT) = 0$. Therefore, $|P(nT)|= 1$, where $n$ is a positive integer. So the system will exactly recover to its initial state after a period $T$. Using Eq. \ref{eq-first-order-two-level}, we can compute the recover probability
$P(t)$, which is shown in Fig. \ref{fig-fig2}(a) - (b) with blue dotted lines. The agreement with the exact many-body solution is excellent. Thus by controlling this ratio $\beta$, our results can be used to realize the multifrequency resonance state with different periods. This period is determined solely by $\beta$, but independent of the other parameters, such as non-integrability terms and time-independent disorders, thus the period has the same robustness as that discussed in Ref. \cite{PhysRevA.98.052324}.

This projection yields a new picture to understand the many-body dynamics, beyond the discussion in the introduction. While the precession in classical mechanics has found wide applications in various fields based on Rabi oscillation,
the nutation dynamics is rarely discussed \cite{mims1972envelope, boscaino1986double, bottcher2012significance}.
We show that the commensurate dynamics between them can be used for the multifrequency resonance phase. Intriguingly, this result shows that the multifrequency resonance may be realized even with classical systems and the two-level systems. The latter setup can be immediately implemented using the architectures for quantum simulation, such as superconducting qubits \cite{PhysRevLett.96.187002,PhysRevLett.101.017003,PhysRevB.81.024520,PhysRevB.83.180507}, ultracold atoms \cite{PhysRevA.95.013827}, quantum dots \cite{cao2013ultrafast,deng2015charge}, color centers \cite{PhysRevA.82.033839,Jiang267,Neumann1326,PhysRevLett.111.067601}, trapped ions \cite{schneider2012experimental,cui2016experimental} and superconducting qubits \cite{tan2018topological,gong2016simulating}.

\begin{figure}
\centering
\includegraphics[width=0.45\textwidth]{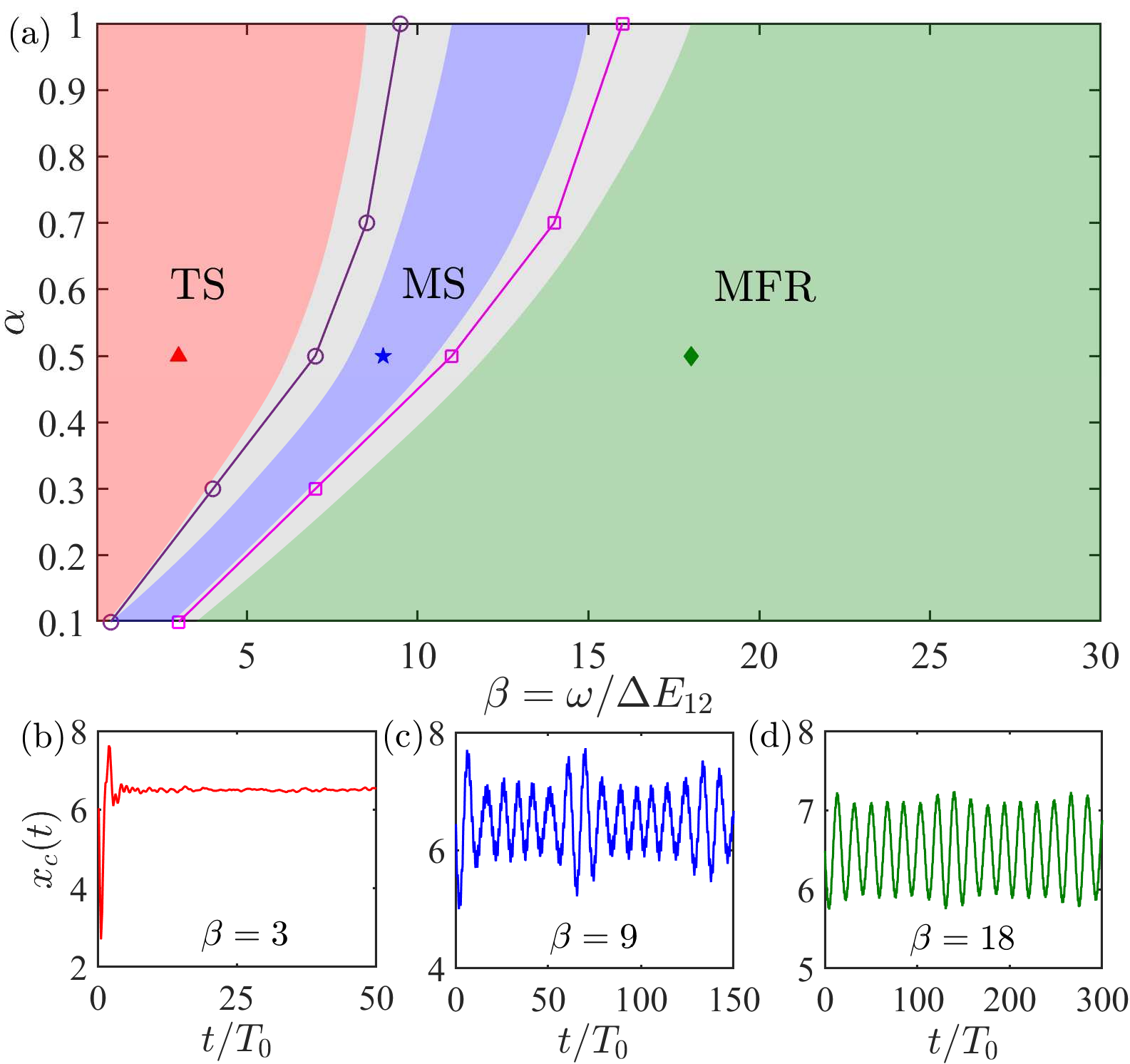}
	\caption{(a) Phase diagram for $L=12$ with ED ($U=0.8$, $N=5$). TS, MS and MFR correspond to thermalized state, mixing state and multifrequency resonance state, respectively. (b) - (d) There examples for these phases for $\alpha = 0.5$ for the three symbols in (a). The grey regimes show the uncertainty of the phase boundaries.}
\label{fig-fig6}
\end{figure}

{\it Phase diagram.} We finally address the phase diagram for the searching of the multifrequency resonance phase. From the above analysis the modulation frequency matters. This modulation frequency has a number 
of important consequences to the dynamics. In the extremely low modulation limit, the system follows adiabatically the dynamics of the Hamiltonian. In the high-frequency
limit, that is, $\omega \gg \text{max}(\epsilon_n) - \text{min}(\epsilon_n)$, the effective dynamics can be well described by the effective Floquet Hamiltonian defined by $U(T_0) = \exp(-iH_\text{F} T_0)$ \cite{blanes2009magnus,goldman2014periodically}. To the third-order approximation \cite{eckardt2017colloquium,eckardt2015high}, $H_\text{F} = -J_\text{F} \sum_i (b_i^\dagger b_{i+1} +h.c.) + \frac{U}{2}\sum_i n_i(n_i-1)$, where 
$J_\text{F} = (1-\frac{5\alpha^2}{12\omega^2})J$. The latter case was widely used in literature for the searching of novel phases, including the  spin-orbit coupling \cite{jackeli2009mott,liu2011low,galitski2013spin} in lattice models as well as some exotic topological phases \cite{liu2014realization, neupane2014observation,li2015exotic}. We are mainly interested in the physics between these two limits. 
In Fig. \ref{fig-fig6}, we show that in the relatively low modulation limit, the wave function is thermalized. With the increasing of modulation frequency, it enters the mixed phase, which exhibits non-regular oscillation. From its Fourier spectroscopy, one may see multiply modulation frequencies due to the coupling of a lot of low-lying eigenstates.
In this way, the period oscillation is hard to be developed in the experimental accessible time. With the further increase of modulation frequency, the dynamics will finally be restricted to the lowest two states, in which almost perfect oscillation can be found. The crossover between these three cases is somewhat smooth, probably due to the finite size effect. We also find the on-set of the multifrequency resonance depends strongly on the inclination $\alpha$. The larger this inclination is, the larger the commensurate ratio is required to be. 

To conclude, we generalize the idea of multifrequency resonance from the single-particle physics to the many-body physics using a quantum seesaw based on an interacting Bose-Hubbard model. This phase is realized by commensurate between precession dynamics and nutation dynamics in the many-body system. In this model, the periodic driving field is used to restrict the dynamics of the wave function to the lowest two energy levels, which may also be realized in the other modulation systems. In this sense, this behavior should be quite general in a lot of periodic driving many-body systems before fully entering the effective Floquet Hamiltonian regime. This kind of dynamics is robust against non-integrability interactions. 

\begin{acknowledgments}
This work is supported by the National Key Research and Development Program in China (Grants No. 2017YFA0304504 and No. 2017YFA0304103) and the National Natural Science Foundation of China (NSFC) with No. 11774328.                                                          
\end{acknowledgments}


%

\end{document}